\documentclass{article}
\usepackage[a4paper,margin=1in]{geometry}
\usepackage{amsmath,amssymb}
\usepackage{graphicx}
\usepackage{bm}
\usepackage{siunitx}
\usepackage{hyperref}
\usepackage{xcolor}
\usepackage{cite}
\usepackage{setspace}
\title{\bf First-Principles Investigation of Electron--Phonon Coupling and Intrinsic Two-Gap Superconductivity in Hexagonal BAs$_3$ Monolayer}

\author{
Jakkapat Seeyangnok$^{1a}$,
Udomsilp Pinsook$^{1b}$\\[1ex]
$^{1}$Department of Physics, Faculty of Science, \\
Chulalongkorn University, Bangkok, Thailand\\[1ex]
\texttt{$^{a}$Jakkapat.Se@chula.ac.th}\\
\texttt{$^{b}$Udomsilp.P@chula.ac.th}\\
}

\date{} 

\begin{document}

\maketitle

\begin{abstract}
Two-dimensional superconductors with multiband electronic structures provide an ideal platform for exploring anisotropic and multigap superconductivity in the reduced-dimensionality limit. Here, we investigate the structural, electronic, vibrational, and superconducting properties of a hexagonal BAs$_3$ monolayer using first-principles calculations combined with density functional perturbation theory and fully anisotropic Migdal--Eliashberg theory. The optimized structure is found to be dynamically and thermally stable, as confirmed by phonon calculations and ab initio molecular dynamics simulations. Electronic structure calculations reveal an intrinsic metallic state with multiple bands crossing the Fermi level and several disconnected Fermi-surface sheets derived primarily from hybridized B-$p$ and As-$p$ orbitals. The electron--phonon interaction is dominated by low-frequency As-derived phonon modes, yielding a total electron--phonon coupling constant of $\lambda=0.75$. Solving the anisotropic Eliashberg equations predicts a superconducting critical temperature of $T_c=3.4$ K. The momentum-resolved superconducting gap exhibits a pronounced two-gap character with gap magnitudes of $\Delta_1=0.75$ meV and $\Delta_2=0.51$ meV at $T=1$ K. The superconducting gaps remain finite over the entire Fermi surface, demonstrating a fully gapped nodeless superconducting state. Analysis of the momentum-dependent electron--phonon coupling reveals that the two-gap superconductivity originates from sheet-dependent pairing interactions associated with distinct Fermi-surface sheets. Our results establish monolayer BAs$_3$ as an intrinsic anisotropic two-gap superconductor and expand the growing family of boron-based two-dimensional superconductors.
\end{abstract}


\section{Introduction}
Low-dimensional materials have emerged as an important platform for investigating emergent quantum phenomena and developing next-generation electronic technologies. Since the isolation of graphene, a diverse family of atomically thin materials has been realized, including transition-metal dichalcogenides, elemental monolayers, MXenes, and various binary compounds, exhibiting a broad range of electronic, magnetic, optical, and superconducting properties~\cite{novoselov2004electric,geim2013van,novoselov20162d,xu2013graphene,manzeli20172d}. Among these, two-dimensional superconductors have attracted particular interest because reduced dimensionality can substantially modify electron--phonon interactions, enhance many-body effects, and stabilize novel quantum phases that are absent in bulk materials~\cite{uchihashi2017two,saito2016superconductivity,xi2016ising}.

Within the framework of Migdal--Eliashberg theory, superconductivity in low-dimensional materials is governed by the interplay between electronic states near the Fermi level and lattice vibrations~\cite{migdal1958interaction,eliashberg1960interactions,giustino2017electron}. This mechanism has been demonstrated in numerous 2D systems, including doped graphene~\cite{profeta2012phonon}, borophene~\cite{gao2017prediction,penev2012polymorphism}, transition-metal dichalcogenides~\cite{ge2015superconductivity,ugeda2016characterization}, MXenes, and hydrogenated layered materials. In systems possessing multiple Fermi-surface sheets, superconductivity often exhibits pronounced anisotropy and multiband characteristics, as exemplified by the prototypical superconductor MgB$_2$~\cite{nagamatsu2001superconductivity,souma2003origin,liu2001beyond}. Such multiband electronic structures can generate distinct superconducting gaps on different Fermi-surface sheets, giving rise to two-gap superconductivity. Recent studies have identified this behavior in a growing number of atomically thin materials, including $n$-doped graphene~\cite{margine2014two}, AlB$_2$-based thin films~\cite{zhao2019two}, trilayer LiB$_2$C$_2$~\cite{gao2020strong}, monolayer LiBC~\cite{modak2021prediction}, GaInSLi~\cite{seeyang2025phase_japtwpgap}, MoSLi~\cite{xie2024strong}, MoSeLi~\cite{seeyangnok2026tunable}, MoSH~\cite{liu2022two}, and hydrogenated borides (MgB$_4$H, CaB$_4$H, and AlB$_4$H)~\cite{seeyangnok2026stability}, highlighting the growing prevalence of intrinsic multigap superconductivity in low-dimensional systems.

The discovery of superconductivity in MgB$_2$ with a transition temperature of 39 K established metal borides as a prominent family of conventional phonon-mediated superconductors~\cite{nagamatsu2001superconductivity}. Subsequent theoretical and experimental studies demonstrated that strong electron--phonon coupling can persist down to the two-dimensional limit. For example, free-standing MgB$_2$ monolayers and ultrathin films retain superconductivity with transition temperatures approaching those of the bulk compound~\cite{bekaert2017free,cheng2018fabrication}. Motivated by these findings, numerous two-dimensional borides have been proposed as promising superconducting candidates, including MB$_4$ compounds ($M =$ Be, Mg, Ca, Sc, and Al), with predicted superconducting transition temperatures reaching 36 K~\cite{sevik2022high}. More recently, chemical functionalization, particularly hydrogenation, has emerged as an effective route for tuning the electronic structure, enhancing electron--phonon coupling, and stabilizing superconductivity in low-dimensional boron-based materials~\cite{seeyangnok2026stability}. Representative examples include Ti$_2$CSH with a predicted $T_c$ exceeding 20 K~\cite{jseeyang_ti2csh} and a series of hydrogenated tungsten-based Janus monolayers exhibiting superconducting transition temperatures around 10--15 K~\cite{seeyangnok2024superconductivity,seeyangnok2024superconductivitywseh,qiao2024prediction,gan2024hydrogenation,fu2024superconductivity}. These systems further reveal a rich interplay between superconductivity and competing quantum phases, including magnetism~\cite{seeyangnok2025competition} and charge-density-wave (CDW) order~\cite{seeyangnok2024superconductivity,seeyangnok2024superconductivitywseh,qiao2024prediction,gan2024hydrogenation,fu2024superconductivity,seeyangnok2026triangular,seeyangnok2026electron,seeyangnok2026enhanced}. In many cases, exceptionally strong electron--phonon coupling drives lattice instabilities and CDW formation, which may either compete with or coexist alongside superconductivity. These studies highlight the remarkable versatility of boron-containing and chemically functionalized two-dimensional materials as a fertile platform for investigating phonon-mediated superconductivity and its interplay with other emergent collective phenomena.

Beyond pure borides, boron--pnictogen compounds have recently attracted increasing attention owing to their diverse bonding environments and tunable electronic structures~\cite{oganov2009ionic,zhang2017two,carvalho2016phosphorene}. Boron forms electron-deficient covalent networks characterized by multicenter bonding, whereas pnictogen atoms introduce directional bonding and strong orbital hybridization~\cite{pancharatna2022anatomy,liu2014phosphorene,xia2014rediscovering}. The interplay between these distinct bonding motifs can produce unusual electronic structures and potentially favorable conditions for superconductivity. In particular, arsenic-based compounds are expected to exhibit stronger spin--orbit coupling and more spatially extended orbitals than their phosphorus counterparts, which may significantly influence both the electronic structure and electron--phonon interactions.

Recently, a hexagonal BAs$_3$ monolayer has been proposed as a stable two-dimensional material with promising electrochemical performance for sodium-ion battery applications~\cite{vu2026bp}. In a closely related boron--pnictogen system, BP$_3$ was predicted to host strong-coupling multiband superconductivity with a pronounced two-gap structure arising from multiple Fermi-surface sheets~\cite{seeyangnok2026strong_bp3}. The structural similarity between BP$_3$ and BAs$_3$ naturally raises the question of whether analogous superconducting behavior can emerge in arsenic-based compounds. However, the electronic, vibrational, and superconducting properties of BAs$_3$ remain largely unexplored. Given the strong hybridization between B-$p$ and As-$p$ orbitals and the growing evidence for superconductivity in boron-based materials, BAs$_3$ represents a promising candidate for investigating phonon-mediated multigap superconductivity in two dimensions.

In this work, we perform a comprehensive first-principles investigation of the structural, electronic, vibrational, and superconducting properties of monolayer BAs$_3$. We demonstrate that the material is dynamically and thermally stable and exhibits a metallic electronic structure characterized by multiple bands crossing the Fermi level. By combining density functional perturbation theory with anisotropic Migdal--Eliashberg calculations, we show that BAs$_3$ hosts a phonon-mediated superconducting state driven by substantial electron--phonon coupling. The superconducting gap exhibits pronounced anisotropy and a clear two-gap character associated with the multiband Fermi surface, revealing the important role of orbital hybridization in determining the superconducting properties. Our results establish BAs$_3$ as a promising member of the growing family of two-gap two-dimensional superconductors and provide new insight into superconductivity in boron--pnictogen materials.

\section{Computational Methods}

First-principles calculations were performed within the framework of density functional theory (DFT) using the \textsc{Quantum ESPRESSO} package~\cite{giannozzi2009quantum}. Exchange--correlation effects were described within the generalized gradient approximation (GGA) using the Perdew--Burke--Ernzerhof (PBE) functional~\cite{perdew1996generalized}. The electron--ion interaction was treated using optimized norm-conserving Vanderbilt (ONCV) pseudopotentials~\cite{hamann2013optimized,schlipf2015optimization}. A plane-wave kinetic-energy cutoff of 80~Ry and a charge-density cutoff of 320~Ry were employed throughout the calculations. Brillouin-zone integrations were carried out using a $12\times12\times1$ Monkhorst--Pack $\mathbf{k}$-point mesh~\cite{monkhorst1976special}, together with Methfessel--Paxton smearing~\cite{methfessel1989high} of width 0.02~Ry.

The atomic structures were fully optimized using the Broyden--Fletcher--Goldfarb--Shanno (BFGS) algorithm~\cite{liu1989limited} until the residual forces on all atoms were smaller than $10^{-5}$~eV/\AA. Vibrational properties and phonon dispersions were computed within density functional perturbation theory (DFPT) using a $6\times6\times1$ $\mathbf{q}$-point mesh.

The superconducting properties were investigated by solving the fully anisotropic Migdal--Eliashberg equations~\cite{migdal1958interaction,eliashberg1960interactions,nambu1960quasi,pinsook2024analytic} as implemented in the EPW package~\cite{noffsinger2010epw,ponce2016epw}. Electron--phonon matrix elements were interpolated from coarse DFT and DFPT grids onto dense momentum meshes using the Wannier--Fourier interpolation technique~\cite{giustino2007electron,giustino2017electron}. The superconducting gap function $\Delta_{n\mathbf{k}}(i\omega_j)$ and mass-renormalization function $Z_{n\mathbf{k}}(i\omega_j)$ were obtained self-consistently on the imaginary-frequency axis according to

\begin{eqnarray}
Z_{n\mathbf{k}}(i\omega_j) &=& 1
+\frac{\pi T}{N(\varepsilon_F)\omega_j}
\sum_{m\mathbf{k}'j'}
\frac{\omega_{j'}}
{\sqrt{\omega_{j'}^{2}+\Delta_{m\mathbf{k}'}^{2}(i\omega_{j'})}}
\nonumber\\
&&\times
\lambda(n\mathbf{k},m\mathbf{k}',\omega_j-\omega_{j'})
\,\delta(\epsilon_{m\mathbf{k}'}-\varepsilon_F),
\end{eqnarray}

\begin{eqnarray}
Z_{n\mathbf{k}}(i\omega_j)\Delta_{n\mathbf{k}}(i\omega_j)
&=&
\frac{\pi T}{N(\varepsilon_F)}
\sum_{m\mathbf{k}'j'}
\frac{\Delta_{m\mathbf{k}'}(i\omega_{j'})}
{\sqrt{\omega_{j'}^{2}+\Delta_{m\mathbf{k}'}^{2}(i\omega_{j'})}}
\delta(\epsilon_{m\mathbf{k}'}-\varepsilon_F)\nonumber\\
&&\times
\left[
\lambda(n\mathbf{k},m\mathbf{k}',\omega_j-\omega_{j'})
-\mu^{*}
\right],
\end{eqnarray}

where $\omega_j=(2j+1)\pi T$ denotes the Matsubara frequencies, $N(\varepsilon_F)$ is the electronic density of states at the Fermi level, and $\mu^{*}$ is the Coulomb pseudopotential, fixed at 0.10.

To achieve converged superconducting properties, dense $\mathbf{k}$- and $\mathbf{q}$-point meshes of $120\times120\times1$ and $60\times60\times1$, respectively, were employed in the EPW calculations. The Dirac delta functions entering the electron--phonon coupling expressions were approximated by Gaussian broadenings of 0.40~eV for electronic states and 0.5~meV for phonon modes. A Fermi-surface broadening of 0.40~eV and a Matsubara frequency cutoff of 1.00~eV were used in solving the anisotropic Migdal--Eliashberg equations.

\section{Results and Discussion}

\subsection{Crystal Structure and Stability}
    \begin{figure}[ht]
        \centering
        \includegraphics[width=14cm]{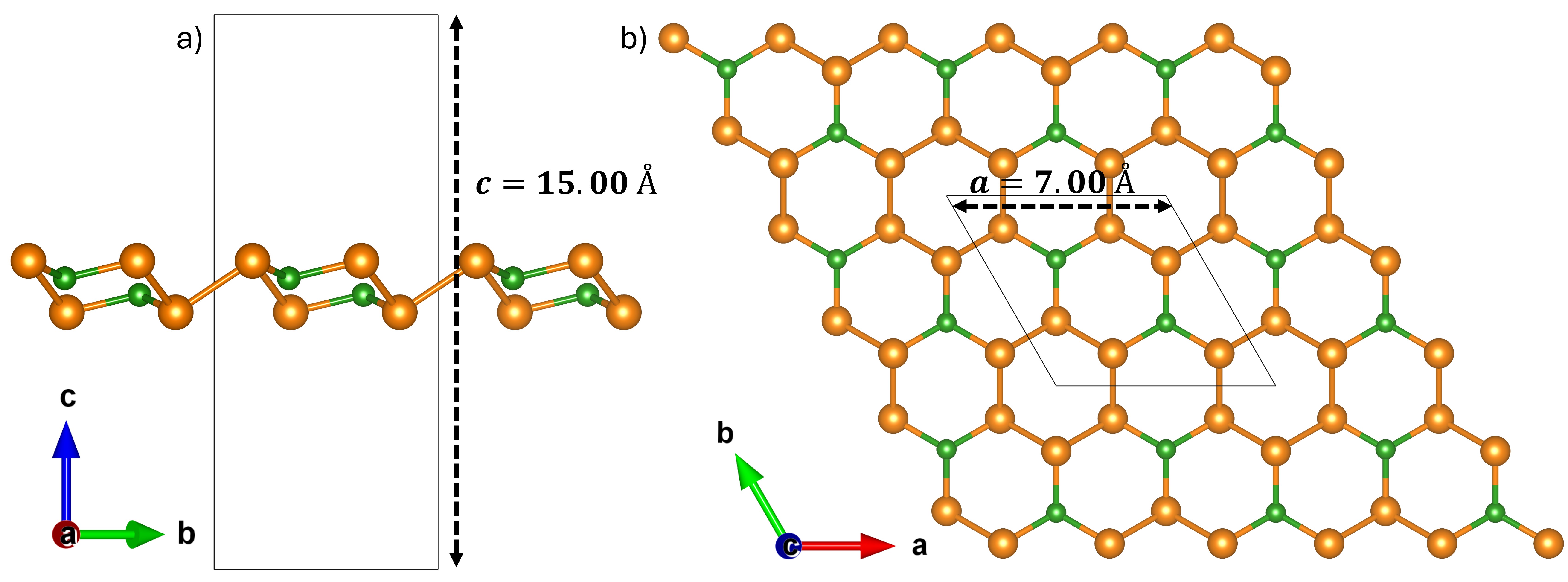}
        \caption{Crystal structure of monolayer BAs$_3$. (a) Side view of the optimized structure showing the buckled boron--arsenic network together with the vacuum region of 15.0~\AA\ introduced along the out-of-plane direction to eliminate interactions between periodic images. (b) Top view of the hexagonal lattice. The primitive unit cell is indicated by the black rhombus, with an optimized lattice constant of $a = 7.00$~\AA. Orange and green spheres represent As and B atoms, respectively.}
        \label{fig:structure}
    \end{figure}

The optimized crystal structure of monolayer BAs$_3$ is shown in Fig.~\ref{fig:structure}. The material crystallizes in a hexagonal lattice consisting of a planar boron framework interconnected by arsenic atoms. The primitive unit cell contains one B atom and three As atoms, corresponding to the chemical composition BAs$_3$. After full structural optimization, the equilibrium lattice constant is found to be $a = 7.00$~\AA. To model the isolated monolayer, a vacuum region of 15.0~\AA\ was introduced along the out-of-plane direction, which is sufficient to suppress spurious interactions between neighboring periodic images.

    \begin{figure}[h!]
        \centering
        \includegraphics[width=14cm]{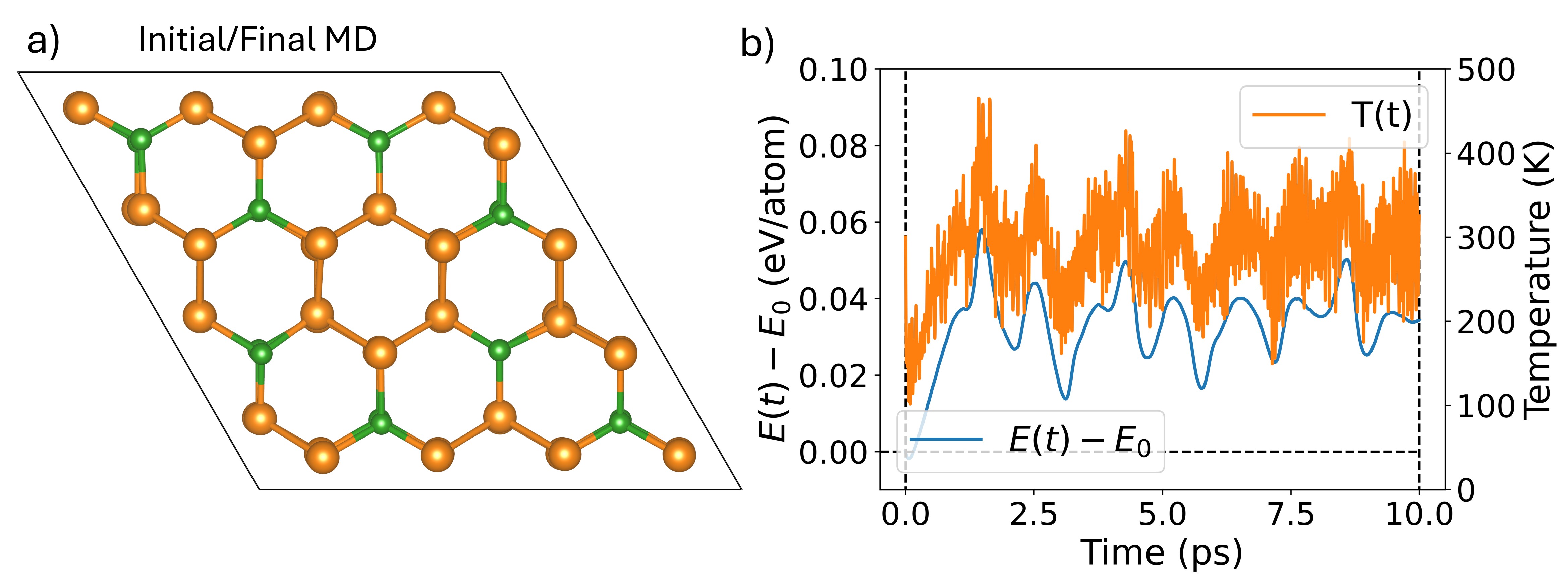}
        \caption{(a) Superposition of the initial and final atomic configurations obtained from a 10 ps AIMD simulation at 300 K. The strong overlap between the two structures indicates only minor thermal displacements and confirms the structural stability of the BAs$_3$ monolayer. Orange and green spheres represent As and B atoms, respectively.(b) Time evolution of the total energy difference per atom, $E(t)-E_0$, (left axis) and instantaneous temperature $T(t)$ (right axis) during the AIMD simulation. The temperature fluctuates around the target value of 300 K, while the total energy remains bounded without noticeable drift, demonstrating the thermal stability of the monolayer.}
        \label{fig:aimd}
    \end{figure}

    To further evaluate the thermal stability of the BAs$_3$ monolayer, ab initio molecular dynamics (AIMD) simulations were performed within the canonical (NVT) ensemble at 300 K using a time step of 1 fs for a total simulation time of 10 ps. The evolution of the total energy and instantaneous temperature is shown in Fig.~\ref{fig:aimd}(b). The temperature fluctuates around the target value of 300 K throughout the simulation, with the relatively large fluctuations originating from the finite size of the simulation cell containing only 32 atoms. Such behavior is typical for NVT simulations of low-dimensional materials and does not indicate any thermodynamic instability.

    The total energy difference, $E(t)-E_0$, remains bounded during the entire simulation and exhibits only periodic oscillations associated with thermal vibrations. Importantly, no continuous energy drift is observed, indicating good numerical stability and the absence of any irreversible structural transformation. The energy fluctuations remain below approximately 0.06 eV/atom, further suggesting that the system remains close to thermal equilibrium throughout the simulation.

    Figure~\ref{fig:aimd}(a) compares the initial and final atomic  configurations by superimposing the two structures. The nearly complete overlap between the initial and final geometries demonstrates that only small thermal displacements occur during the simulation. No bond breaking, reconstruction, clustering, or significant out-of-plane distortion is observed, and the characteristic hexagonal network remains intact. These observations confirm that the BAs$_3$ monolayer preserves its structural integrity under ambient conditions.

\subsection{Electronic Structure, and Bonding Characteristics}
    \begin{figure}[h!]
        \centering
        \includegraphics[width=14cm]{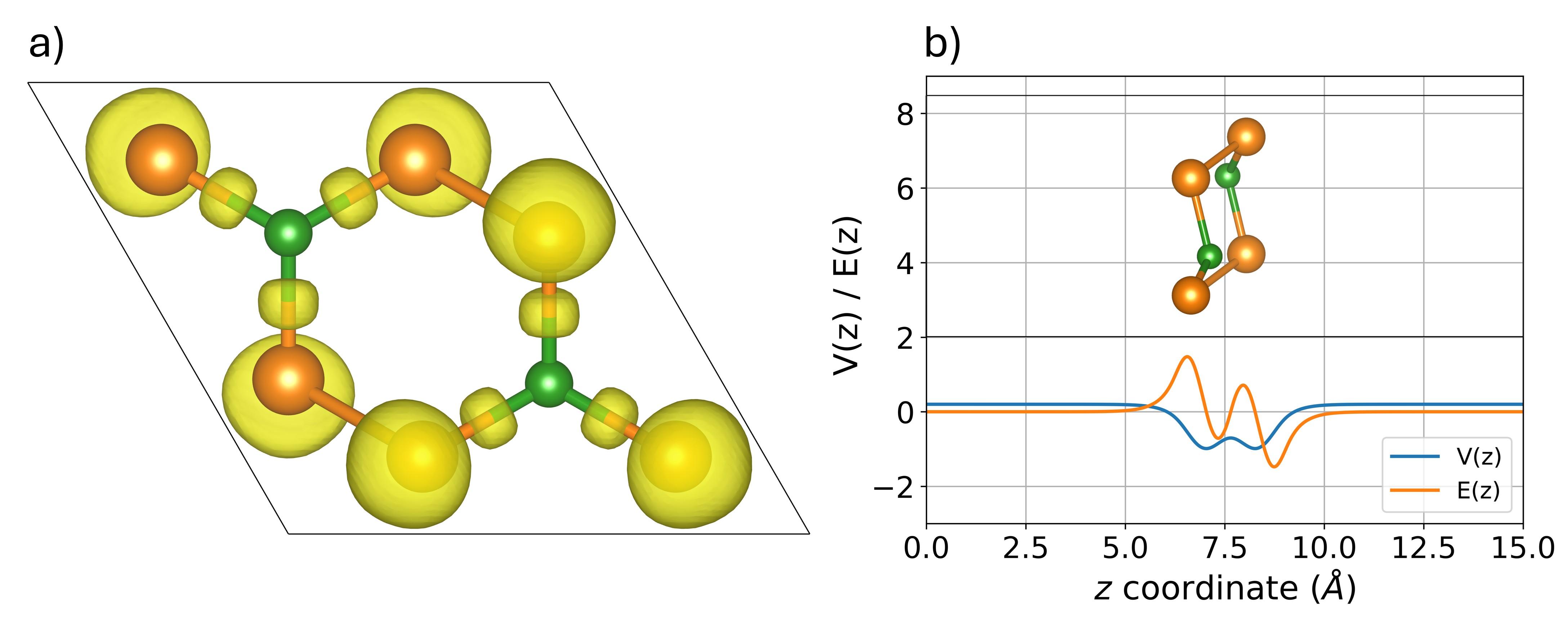}
    \caption{Bonding character and electrostatic profile of monolayer BAs$_3$. (a) Electron localization function (ELF) of monolayer BAs$_3$. The pronounced ELF localization along the B--As bonds indicates strong directional covalent bonding arising from B-$p$ and As-$p$ orbital hybridization. (b) Plane-averaged electrostatic potential $V(z)$ and corresponding electric field $E(z)$ along the out-of-plane direction. The asymmetric potential profile around the atomic layer reflects the polar nature of the BAs$_3$ monolayer, while both $V(z)$ and $E(z)$ become nearly constant in the vacuum region.}
    \label{fig:elf_bader}
    \end{figure}

    Figure~\ref{fig:elf_bader}(a) shows the electron localization function (ELF) of monolayer BAs$_3$. Pronounced ELF maxima are observed along the B--As bonds, indicating substantial electron localization between neighboring atoms. This behavior is characteristic of directional covalent bonding and reflects strong orbital hybridization between B-$p$ and As-$p$ states. The continuous ELF distribution connecting adjacent atoms demonstrates that the bonding is predominantly covalent rather than purely ionic. 

    To quantify the charge redistribution, Bader charge analysis was performed. The results indicate a net transfer of electronic charge from As to B atoms, with each B atom gaining approximately 4.2 electrons and each As atom losing about 1.4 electrons. Although charge transfer is present, the pronounced electron localization along the B--As bonds revealed by the ELF demonstrates that the bonding retains a strong covalent character. Therefore, the bonding in BAs$_3$ can be described as a mixed ionic--covalent interaction arising from significant B-$p$/As-$p$ hybridization.

\subsection{Electronic Structure and Fermi Surface}
    \begin{figure}[h!]
        \centering
        \includegraphics[width=13cm]{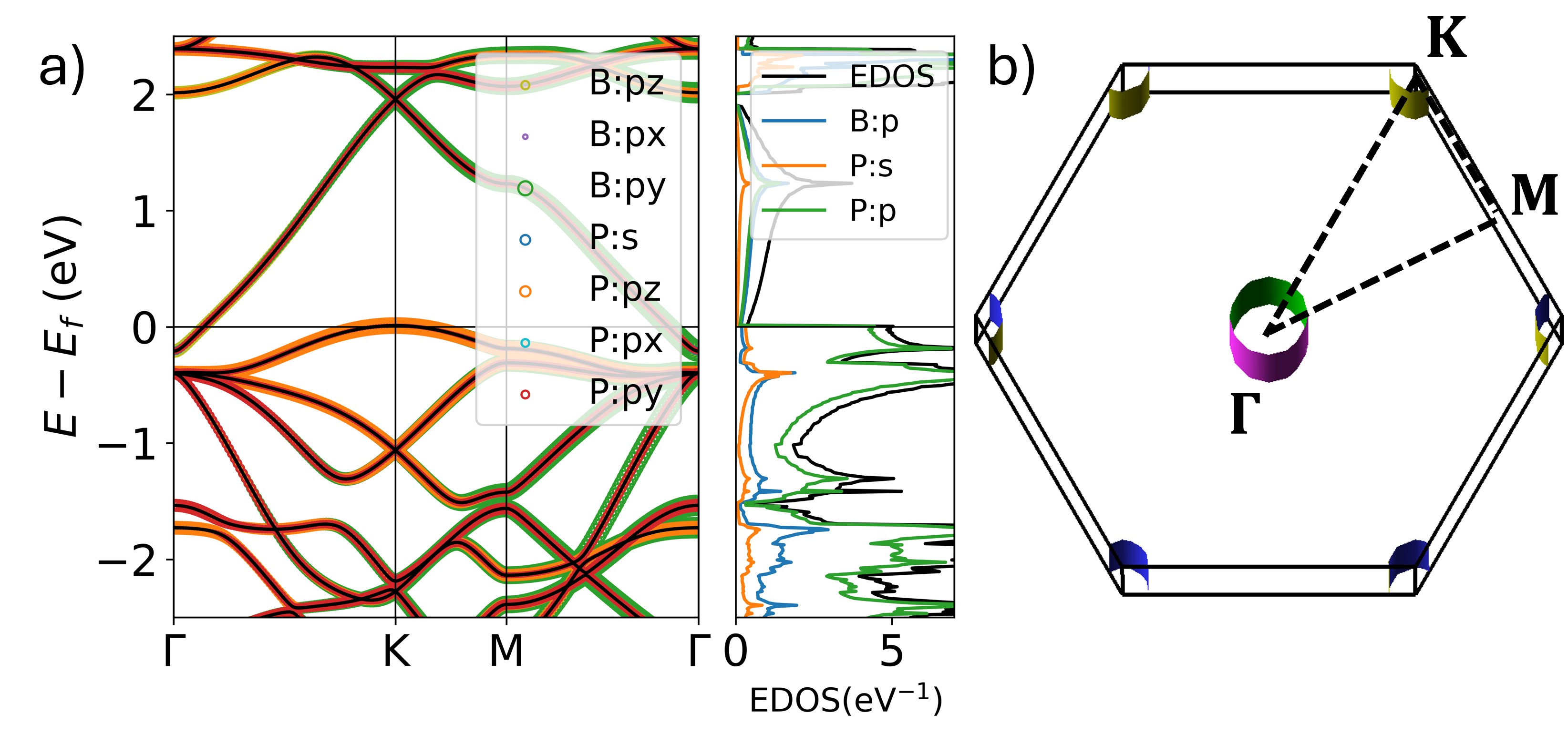}
        \caption{ Electronic structure of monolayer BAs$_3$. (a) Orbital-projected electronic band structure and electronic density of states (EDOS). The Fermi level is set to zero energy. The size of the colored markers is proportional to the orbital contribution of the corresponding atomic states. The states near the Fermi level are dominated by B-$p$ and As-$p$ orbitals, indicating strong orbital hybridization. The EDOS confirms the metallic character of the system with a finite density of states at the Fermi level. (b) Calculated Fermi surface of monolayer BAs$_3$. Multiple disconnected Fermi-surface sheets are present, including a cylindrical pocket centered at the $\Gamma$ point and additional pockets located near the Brillouin-zone boundaries. The coexistence of several Fermi-surface sheets suggests multiband electronic behavior and provides a favorable condition for anisotropic and multigap superconductivity.}
    \label{fig:bands_fs}
    \end{figure}
    
The orbital-projected electronic band structure and electronic density of states (EDOS) of monolayer BAs$_3$ are presented in Fig.~\ref{fig:bands_fs}(a). The calculated band structure reveals several bands crossing the Fermi level, demonstrating the intrinsic metallic nature of the material. The electronic states in the vicinity of the Fermi level are primarily derived from B-$p$ and As-$p$ orbitals. In particular, the As-$p$ orbitals provide the dominant contribution to the density of states around $E_F$, while significant B-$p$ character is also present. The strong overlap between these orbital components indicates substantial hybridization between B-$p$ and As-$p$ states, consistent with the covalent bonding characteristics discussed previously. Such hybridization plays a central role in determining both the transport and superconducting properties of the material.

The orbital-resolved band structure and electronic density of states shown in Fig.~\ref{fig:bands}(a) demonstrate that the low-energy electronic structure of monolayer BAs$_3$ is dominated by hybridized B-$p$ and As-$p$ states. The bands crossing the Fermi level possess predominantly $p_z$ character, indicating that the out-of-plane orbitals govern the electronic states participating in superconductivity. Analysis of the Fermi surface reveals a clear orbital separation between the different sheets: the central pocket around the $\Gamma$ point originates mainly from As-$p_z$ orbitals, whereas the pockets near the zone edges and corners around K are primarily composed of B-$p_z$ states. This orbital-selective Fermi-surface topology is particularly important because it enables distinct electron--phonon coupling strengths on different sheets. Consequently, the As-$p_z$- and B-$p_z$-derived Fermi surfaces contribute differently to the superconducting pairing interaction, providing a natural microscopic origin for the anisotropic two-gap superconductivity obtained from the anisotropic Migdal--Eliashberg calculations.

\subsection{Phonon Spectrum and Electron--Phonon Coupling}
\begin{figure}[h!]
    \centering
    \includegraphics[width=11cm]{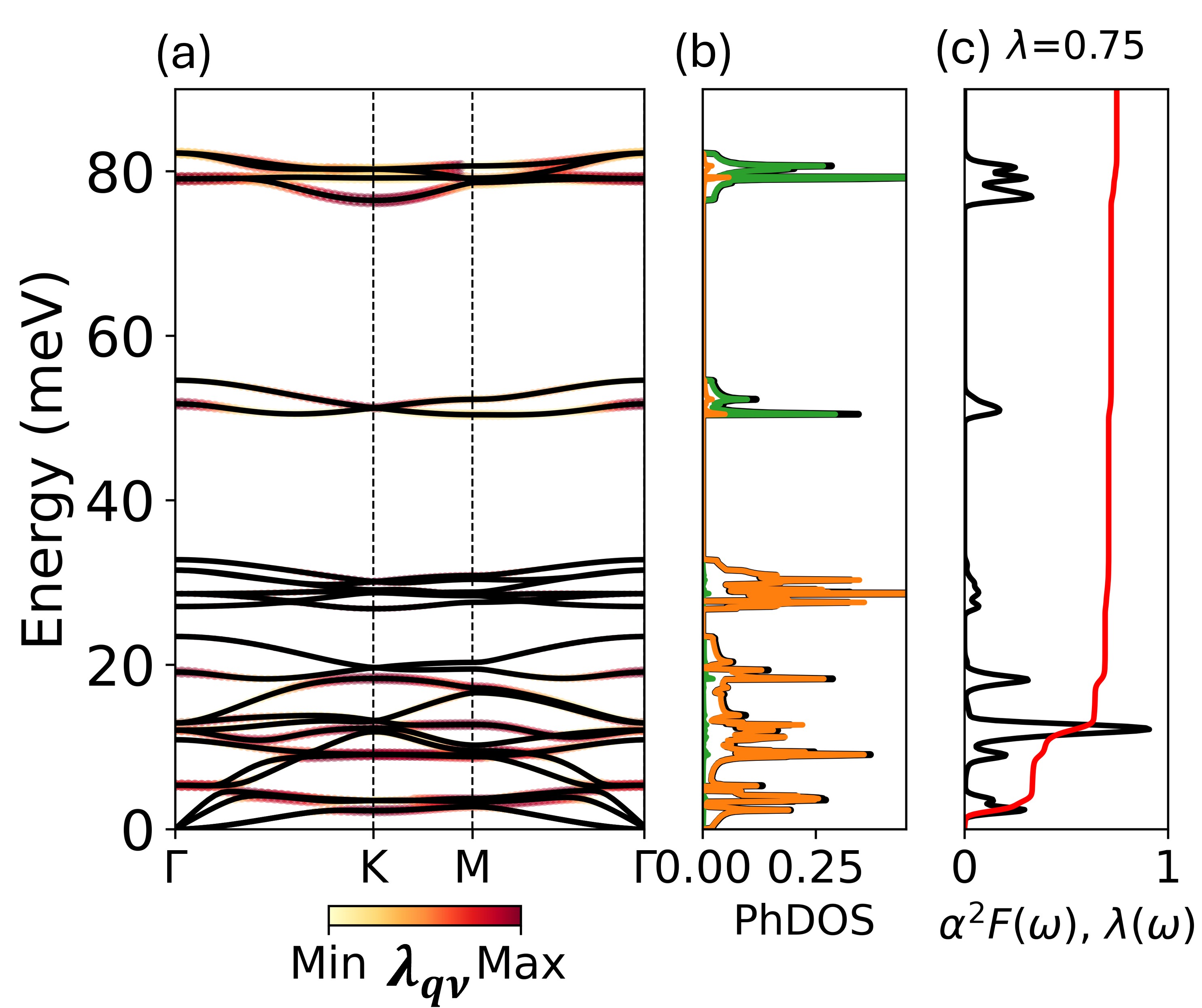}
    \caption{Phonon and electron--phonon coupling (EPC) properties of monolayer BAs$_3$. (a) Phonon dispersion relations along the high-symmetry path in the Brillouin zone. The color scale represents the mode-resolved EPC strength $\lambda_{\mathbf{q}\nu}$, with darker red colors indicating stronger coupling. (b) Total phonon density of states (PhDOS, black line) together with the projected phonon density of states of B (green) and As (orange) atoms. The low- and intermediate-frequency vibrations are predominantly associated with As atoms, whereas the high-frequency optical modes mainly originate from the lighter B atoms. (c) Eliashberg spectral function $\alpha^2F(\omega)$ (black line) and cumulative electron--phonon coupling strength $\lambda(\omega)$ (red line). The total EPC constant is $\lambda=0.75$.}
    \label{fig:epc}
    \end{figure}

To assess the lattice dynamics and superconducting properties of monolayer BAs$_3$, we calculated the phonon spectrum and electron--phonon coupling (EPC) properties within density functional perturbation theory. The phonon dispersion relations are shown in Fig.~\ref{fig:epc}(a). All phonon branches exhibit positive frequencies throughout the Brillouin zone, demonstrating that the optimized structure is dynamically stable. The absence of soft modes or imaginary frequencies indicates that the monolayer is robust against structural distortions and lattice instabilities. The color map superimposed on the phonon dispersion represents the mode-resolved EPC strength $\lambda_{\mathbf{q}\nu}$. Enhanced EPC is primarily concentrated in the low-frequency acoustic and low-lying optical branches below approximately 20 meV. In contrast, the high-frequency optical modes contribute less significantly to the total EPC. This behavior suggests that superconductivity in BAs$_3$ is mainly driven by low-energy lattice vibrations.

Further insight into the vibrational properties is obtained from the projected phonon density of states shown in Fig.~\ref{fig:epc}(b). Owing to the large mass difference between B and As atoms, the phonon spectrum is naturally divided into distinct frequency regions. The low- and intermediate-frequency modes are dominated by vibrations of the heavier As atoms, whereas the high-frequency optical modes mainly originate from the lighter B atoms. Such separation of vibrational character is typical of binary compounds with substantial mass contrast.

The Eliashberg spectral function $\alpha^2F(\omega)$ and cumulative EPC parameter $\lambda(\omega)$ are presented in Fig.~\ref{fig:epc}(c). Pronounced peaks in $\alpha^2F(\omega)$ occur in the low-frequency region, indicating strong coupling between electrons and As-dominated phonon modes. Correspondingly, the cumulative EPC rapidly increases below 20 meV and then gradually saturates at higher energies. The total EPC constant is found to be $\lambda=0.75$, placing BAs$_3$ in the intermediate-coupling regime of phonon-mediated superconductors.

\subsection{Anisotropic Superconductivity}
\begin{figure}[ht]
    \centering
    \includegraphics[width=14cm]{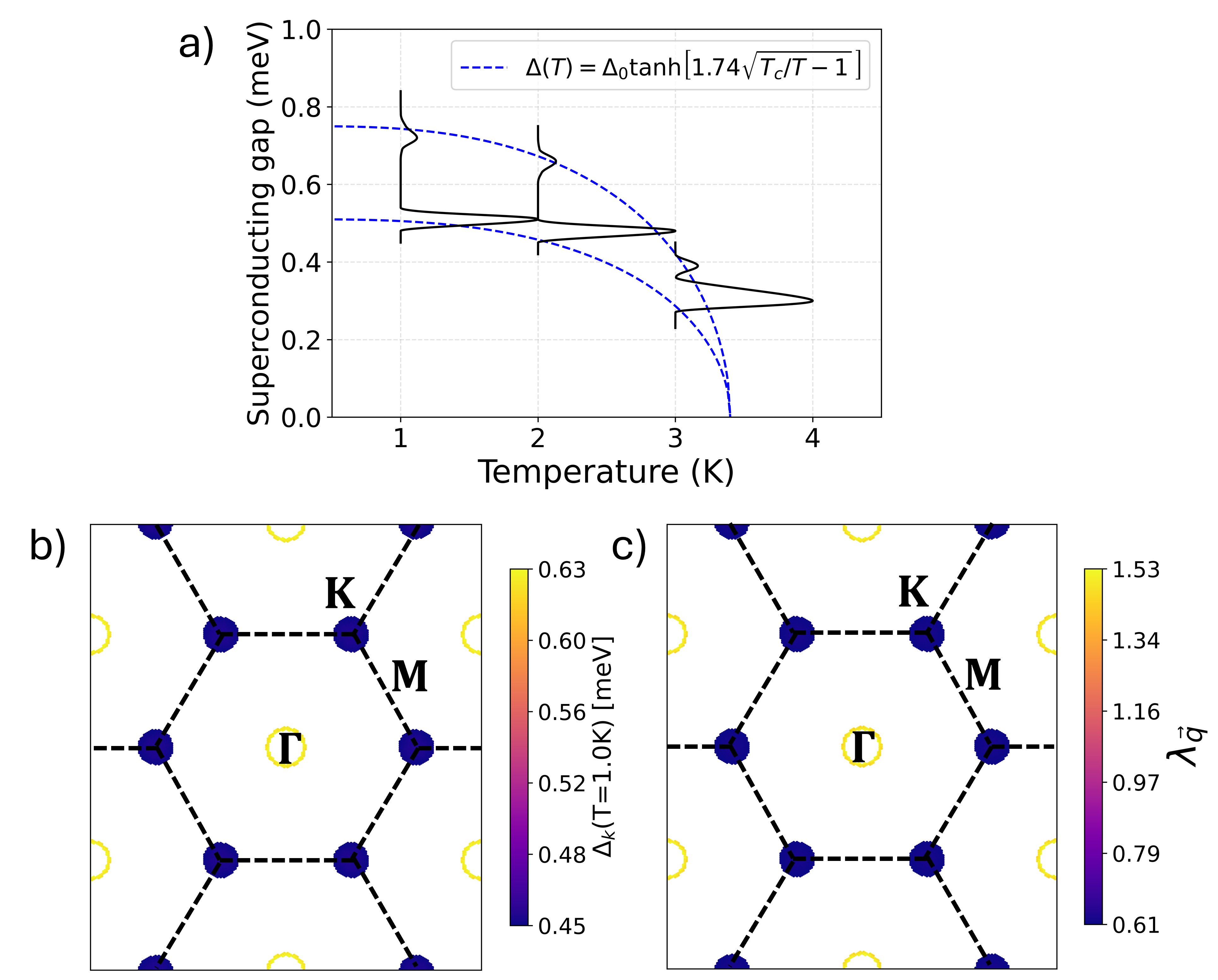}
    \caption{(a) Temperature dependence of the superconducting gaps obtained from anisotropic Eliashberg calculations for the hexagonal BAs$_3$ monolayer. The black curves represent the gap distributions at selected temperatures, while the blue dashed lines denote the isotropic BCS gap function $\Delta(T)=\Delta_0\tanh[1.74\sqrt{T_c/T-1}]$ using $\Delta_0=0.75$ and $0.51$ meV with $T_c=3.4$ K. (b) Momentum-resolved superconducting gap $\Delta_{\mathbf{k}}$ on the Fermi surface at $T=1.0$ K. The color scale represents the magnitude of the superconducting gap, revealing two distinct gap amplitudes associated with different Fermi-surface sheets. (c) Momentum-resolved electron--phonon coupling strength $\lambda_{\mathbf{k}}$ on the Fermi surface.}
    \label{fig:gap}
    \end{figure}

To further elucidate the superconducting state of the hexagonal BAs$_3$ monolayer, we performed fully anisotropic Eliashberg calculations. Figure~\ref{fig:gap}(a) shows the temperature evolution of gap distributions. At low temperature ($T=1$ K), the superconducting state is characterized by two distinct gap magnitudes of approximately $\Delta_1=0.75$ meV and $\Delta_2=0.51$ meV. Both gaps decrease continuously with increasing temperature and vanish at the superconducting critical temperature $T_c=3.4$ K. The temperature dependence of the two gaps follows the conventional BCS-like behavior,
\begin{equation}
\Delta(T)=\Delta_0\tanh\!\left[1.74\sqrt{\frac{T_c}{T}-1}\right],
\end{equation}
indicating that superconductivity remains phonon-mediated despite the pronounced gap anisotropy.

The momentum-resolved gap distribution shown in Fig.~\ref{fig:gap}(b) provides direct evidence for multigap superconductivity. The superconducting gap remains finite over the entire Fermi surface, demonstrating a fully gapped and nodeless superconducting state. However, the gap magnitude exhibits a clear sheet dependence, resulting in two well-separated superconducting gaps. This behavior originates from the presence of two distinct Fermi-surface sheets derived primarily from the B-$p_z$ and As-$p_z$ orbitals. The larger gap is concentrated on one Fermi-surface sheet, whereas the smaller gap develops on the other, reflecting the different strengths of electron--phonon coupling experienced by electrons on each sheet. 

The origin of the gap anisotropy can be understood from the momentum-resolved electron--phonon coupling parameter $\lambda_{\mathbf{k}}$ shown in Fig.~\ref{fig:gap}(c). Regions exhibiting larger superconducting gaps coincide with stronger electron--phonon coupling, while smaller-gap regions correspond to weaker coupling strengths. This strong correlation between $\Delta_{\mathbf{k}}$ and $\lambda_{\mathbf{k}}$ demonstrates that the multigap superconductivity arises from the highly anisotropic distribution of electron--phonon interactions across the Fermi surface.

Therefore, the hexagonal BAs$_3$ monolayer can be classified as a strongly coupled anisotropic two-gap superconductor with a critical temperature of $3.4$ K. The coexistence of two distinct superconducting gaps and their sheet-dependent electron--phonon coupling places BAs$_3$ among a growing family of intrinsic multigap superconductors, including $n$-doped graphene~\cite{margine2014two}, AlB$_2$-based thin films~\cite{zhao2019two}, trilayer LiB$_2$C$_2$~\cite{gao2020strong}, monolayer LiBC~\cite{modak2021prediction}, GaInSLi~\cite{seeyang2025phase_japtwpgap}, MoSLi~\cite{xie2024strong}, MoSeLi~\cite{seeyangnok2026tunable}, MoSH~\cite{liu2022two}, AsP$_3$~\cite{seeyangnok2026strong_bp3}, and hydrogenated borides (MgB$_4$H, CaB$_4$H, and AlB$_4$H)~\cite{seeyangnok2026stability}. These findings establish BAs$_3$ as an intrinsic multigap superconductor in the two-dimensional limit and highlight the important role of orbital-selective electron--phonon coupling in shaping its superconducting properties.

\section*{Conclusions}
In summary, we have performed a comprehensive first-principles investigation of the structural, electronic, vibrational, and superconducting properties of the hexagonal BAs$_3$ monolayer. Structural optimization, phonon calculations, and ab initio molecular dynamics simulations confirm that the material is dynamically and thermally stable. The electronic structure exhibits intrinsic metallicity with multiple bands crossing the Fermi level and several disconnected Fermi-surface sheets originating from hybridized B-$p$ and As-$p$ orbitals.

Density functional perturbation theory reveals that the electron--phonon interaction is primarily mediated by low-frequency As-dominated phonon modes, resulting in a moderate electron--phonon coupling strength of $\lambda=0.75$. Fully anisotropic Migdal--Eliashberg calculations predict a superconducting transition temperature of $T_c=3.4$ K. More importantly, the superconducting state exhibits a pronounced multigap character with two distinct superconducting gaps of approximately 0.75 and 0.51 meV at low temperature. The superconducting gap remains finite throughout the entire Fermi surface, demonstrating a fully gapped nodeless superconducting state.

The momentum-resolved superconducting gap and electron--phonon coupling distributions reveal a strong correlation between the magnitude of the superconducting gap and the local pairing strength on different Fermi-surface sheets. The larger and smaller superconducting gaps originate from distinct Fermi-surface sheets derived predominantly from B-$p_z$ and As-$p_z$ orbitals, respectively, establishing the orbital-selective nature of the superconducting pairing mechanism. Consequently, BAs$_3$ belongs to the growing class of intrinsic two-gap superconductors alongside several recently predicted low-dimensional superconducting materials.

These findings demonstrate that boron--pnictogen compounds provide a promising platform for realizing anisotropic multigap superconductivity in two dimensions. The coexistence of multiple Fermi-surface sheets, orbital-selective electron--phonon coupling, and intrinsic metallicity makes monolayer BAs$_3$ an attractive candidate for future experimental investigations of multiband superconductivity in atomically thin materials.

\section*{Acknowledgments}
	This research project is supported by the Second Century Fund (C2F), Chulalongkorn University (Grant No. C2F PD-2320260067). High-performance computing facility in this Research is funded by Thailand Science research and Innovation Fund Chulalongkorn University (ST690022300001).

\bibliographystyle{unsrt}
\bibliography{references}

@article{novoselov2004electric,
  title={Electric field effect in atomically thin carbon films},
  author={Novoselov, Kostya S and Geim, Andre K and Morozov, Sergei V and Jiang, De-eng and Zhang, Yanshui and Dubonos, Sergey V and Grigorieva, Irina V and Firsov, Alexandr A},
  journal={science},
  volume={306},
  number={5696},
  pages={666--669},
  year={2004},
  publisher={American Association for the Advancement of Science}
}

@article{seeyangnok2026strong_bp3,
  title={Strong Electron-Phonon Coupling and Multiband Superconductivity in Hexagonal BP3 Monolayer},
  author={Seeyangnok, Jakkapat and Pinsook, Udomsilp},
  journal={arXiv preprint arXiv:2604.10026},
  year={2026}
}

@article{manzeli20172d,
  title={2D transition metal dichalcogenides},
  author={Manzeli, Sajedeh and Ovchinnikov, Dmitry and Pasquier, Diego and Yazyev, Oleg V and Kis, Andras},
  journal={Nature Reviews Materials},
  volume={2},
  number={8},
  pages={17033},
  year={2017},
  publisher={Nature Publishing Group}
}

@article{novoselov20162d,
  title={2D materials and van der Waals heterostructures},
  author={Novoselov, K and Mishchenko, Artem and Carvalho, Alexandra and Castro Neto, AH},
  journal={Science},
  volume={353},
  number={6298},
  pages={aac9439},
  year={2016},
  publisher={American Association for the Advancement of Science}
}

@article{xu2013graphene,
  title={Graphene-like two-dimensional materials},
  author={Xu, Mingsheng and Liang, Tao and Shi, Minmin and Chen, Hongzheng},
  journal={Chemical reviews},
  volume={113},
  number={5},
  pages={3766--3798},
  year={2013},
  publisher={ACS Publications}
}

@article{uchihashi2017two,
  title={Two-dimensional superconductors with atomic-scale thickness},
  author={Uchihashi, Takashi},
  journal={Superconductor Science and Technology},
  volume={30},
  number={1},
  pages={013002},
  year={2017},
  publisher={IOP Publishing}
}

@article{saito2016superconductivity,
  title={Superconductivity protected by spin--valley locking in ion-gated MoS 2},
  author={Saito, Yu and Nakamura, Yasuharu and Bahramy, Mohammad Saeed and Kohama, Yoshimitsu and Ye, Jianting and Kasahara, Yuichi and Nakagawa, Yuji and Onga, Masaru and Tokunaga, Masashi and Nojima, Tsutomu and others},
  journal={Nature Physics},
  volume={12},
  number={2},
  pages={144--149},
  year={2016},
  publisher={Nature Publishing Group UK London}
}

@article{xi2016ising,
  title={Ising pairing in superconducting NbSe 2 atomic layers},
  author={Xi, Xiaoxiang and Wang, Zefang and Zhao, Weiwei and Park, Ju-Hyun and Law, Kam Tuen and Berger, Helmuth and Forr{\'o}, L{\'a}szl{\'o} and Shan, Jie and Mak, Kin Fai},
  journal={Nature Physics},
  volume={12},
  number={2},
  pages={139--143},
  year={2016},
  publisher={Nature Publishing Group UK London}
}

@article{migdal1958interaction,
  title={Interaction between electrons and lattice vibrations in a normal metal},
  author={Migdal, AB},
  journal={Sov. Phys. JETP},
  volume={7},
  number={6},
  pages={996--1001},
  year={1958}
}

@article{eliashberg1960interactions,
  title={Interactions between electrons and lattice vibrations in a superconductor},
  author={Eliashberg, GM},
  journal={Sov. Phys. JETP},
  volume={11},
  number={3},
  pages={696--702},
  year={1960}
}

@article{giustino2017electron,
  title={Electron-phonon interactions from first principles},
  author={Giustino, Feliciano},
  journal={Reviews of Modern Physics},
  volume={89},
  number={1},
  pages={015003},
  year={2017},
  publisher={APS}
}

@article{profeta2012phonon,
  title={Phonon-mediated superconductivity in graphene by lithium deposition},
  author={Profeta, Gianni and Calandra, Matteo and Mauri, Francesco},
  journal={Nature physics},
  volume={8},
  number={2},
  pages={131--134},
  year={2012},
  publisher={Nature Publishing Group UK London}
}

@article{gao2017prediction,
  title={Prediction of phonon-mediated superconductivity in borophene},
  author={Gao, Miao and Li, Qi-Zhi and Yan, Xun-Wang and Wang, Jun},
  journal={Physical Review B},
  volume={95},
  number={2},
  pages={024505},
  year={2017},
  publisher={APS}
}

@article{penev2012polymorphism,
  title={Polymorphism of two-dimensional boron},
  author={Penev, Evgeni S and Bhowmick, Somnath and Sadrzadeh, Arta and Yakobson, Boris I},
  journal={Nano letters},
  volume={12},
  number={5},
  pages={2441--2445},
  year={2012},
  publisher={ACS Publications}
}

@article{ge2015superconductivity,
  title={Superconductivity above 100 K in single-layer FeSe films on doped SrTiO 3},
  author={Ge, Jian-Feng and Liu, Zhi-Long and Liu, Canhua and Gao, Chun-Lei and Qian, Dong and Xue, Qi-Kun and Liu, Ying and Jia, Jin-Feng},
  journal={Nature materials},
  volume={14},
  number={3},
  pages={285--289},
  year={2015},
  publisher={Nature Publishing Group UK London}
}

@article{ugeda2016characterization,
  title={Characterization of collective ground states in single-layer NbSe 2},
  author={Ugeda, Miguel M and Bradley, Aaron J and Zhang, Yi and Onishi, Seita and Chen, Yi and Ruan, Wei and Ojeda-Aristizabal, Claudia and Ryu, Hyejin and Edmonds, Mark T and Tsai, Hsin-Zon and others},
  journal={Nature Physics},
  volume={12},
  number={1},
  pages={92--97},
  year={2016},
  publisher={Nature Publishing Group UK London}
}

@article{souma2003origin,
  title={The origin of multiple superconducting gaps in MgB2},
  author={Souma, S and Machida, Y and Sato, T and Takahashi, T and Matsui, H and Wang, S-C and Ding, H and Kaminski, A and Campuzano, JC and Sasaki, S and others},
  journal={Nature},
  volume={423},
  number={6935},
  pages={65--67},
  year={2003},
  publisher={Nature Publishing Group UK London}
}

@article{liu2001beyond,
  title={Beyond Eliashberg superconductivity in MgB 2: anharmonicity, two-phonon scattering, and multiple gaps},
  author={Liu, Amy Y and Mazin, II and Kortus, Jens},
  journal={Physical Review Letters},
  volume={87},
  number={8},
  pages={087005},
  year={2001},
  publisher={APS}
}

@article{nagamatsu2001superconductivity,
  title={Superconductivity at 39 K in Magnesium Diboride},
  author={Nagamatsu, Jun and Nakagawa, Norimasa and Muranaka, Takahiro and Zenitani, Yuji and Akimitsu, Jun},
  journal={Nature},
  volume={410},
  number={6824},
  pages={63--64},
  year={2001},
  publisher={Nature Publishing Group UK London}
}

@article{bekaert2017free,
  title={Free surfaces recast superconductivity in few-monolayer MgB2: Combined first-principles and ARPES demonstration},
  author={Bekaert, J and Bignardi, L and Aperis, Alex and Van Abswoude, P and Mattevi, C and Gorovikov, S and Petaccia, L and Goldoni, A and Partoens, B and Oppeneer, Peter M and others},
  journal={Scientific reports},
  volume={7},
  number={1},
  pages={14458},
  year={2017},
  publisher={Nature Publishing Group UK London}
}

@article{cheng2018fabrication,
  title={Fabrication and Characterization of Superconducting {MgB$_2$} Thin Film on Graphene},
  author={Cheng, Shu-Han and Zhang, Yan and Wang, Hong-Zhang and Li, Yu-Long and Yang, Can and Wang, Yue},
  journal={AIP Advances},
  volume={8},
  number={7},
  year={2018},
  publisher={AIP Publishing}
}

@article{sevik2022high,
  title={High-Temperature Multigap Superconductivity in Two-Dimensional Metal Borides},
  author={Sevik, Cem and Bekaert, Jonas and Petrov, Mikhail and Milo{\v{s}}evi{\'c}, Milorad V},
  journal={Physical Review Materials},
  volume={6},
  number={2},
  pages={024803},
  year={2022},
  publisher={APS}
}

@article{oganov2009ionic,
  title={Ionic high-pressure form of elemental boron},
  author={Oganov, Artem R and Chen, Jiuhua and Gatti, Carlo and Ma, Yanzhang and Ma, Yanming and Glass, Colin W and Liu, Zhenxian and Yu, Tony and Kurakevych, Oleksandr O and Solozhenko, Vladimir L},
  journal={Nature},
  volume={457},
  number={7231},
  pages={863--867},
  year={2009},
  publisher={Nature Publishing Group UK London}
}

@article{zhang2017two,
  title={Two-dimensional boron: structures, properties and applications},
  author={Zhang, Zhuhua and Penev, Evgeni S and Yakobson, Boris I},
  journal={Chemical Society Reviews},
  volume={46},
  number={22},
  pages={6746--6763},
  year={2017},
  publisher={Royal Society of Chemistry}
}

@article{carvalho2016phosphorene,
  title={Phosphorene: from theory to applications},
  author={Carvalho, Alexandra and Wang, Min and Zhu, Xi and Rodin, Aleksandr S and Su, Haibin and Castro Neto, Antonio H},
  journal={Nature Reviews Materials},
  volume={1},
  number={11},
  pages={16061},
  year={2016},
  publisher={Nature Publishing Group}
}

@article{pancharatna2022anatomy,
  title={Anatomy of classical boron--boron bonding: overlap and sp dissonance},
  author={Pancharatna, Pattath D and Dar, Sohail H and Chowdhury, Unmesh D and Balakrishnarajan, Musiri M},
  journal={The Journal of Physical Chemistry A},
  volume={126},
  number={20},
  pages={3219--3228},
  year={2022},
  publisher={ACS Publications}
}

@article{liu2014phosphorene,
  title={Phosphorene: an unexplored 2D semiconductor with a high hole mobility},
  author={Liu, Han and Neal, Adam T and Zhu, Zhen and Luo, Zhe and Xu, Xianfan and Tom{\'a}nek, David and Ye, Peide D},
  journal={ACS nano},
  volume={8},
  number={4},
  pages={4033--4041},
  year={2014},
  publisher={ACS Publications}
}

@article{xia2014rediscovering,
  title={Rediscovering black phosphorus as an anisotropic layered material for optoelectronics and electronics},
  author={Xia, Fengnian and Wang, Han and Jia, Yichen},
  journal={Nature communications},
  volume={5},
  number={1},
  pages={4458},
  year={2014},
  publisher={Nature Publishing Group UK London}
}

@article{vu2026bp,
  title={The BP 3 monolayer as a high-capacity and rapid-diffusion anode for sodium-ion batteries: a first-principles study},
  author={Vu, Tuan V and Hoang, Duc-Quang and Ho, Thi H and Van Chi, Hoang and Pham, Khang D},
  journal={Nanoscale Advances},
  volume={8},
  number={2},
  pages={673--681},
  year={2026},
  publisher={Royal Society of Chemistry}
}

@article{giannozzi2009quantum,
  title={QUANTUM ESPRESSO: A Modular and Open-Source Software Project for Quantum Simulations of Materials},
  author={Giannozzi, Paolo and Baroni, Stefano and Bonini, Nicola and Calandra, Matteo and Car, Roberto and Cavazzoni, Carlo and Ceresoli, Davide and Chiarotti, Guido L and Cococcioni, Matteo and Dabo, Ismaila and others},
  journal={Journal of Physics: Condensed Matter},
  volume={21},
  number={39},
  pages={395502},
  year={2009},
  publisher={IOP Publishing}
}

@article{monkhorst1976special,
  title={Special Points for Brillouin-Zone Integrations},
  author={Monkhorst, Hendrik J and Pack, James D},
  journal={Physical Review B},
  volume={13},
  number={12},
  pages={5188},
  year={1976},
  publisher={APS}
}

@article{hamann2013optimized,
  title={Optimized Norm-Conserving Vanderbilt Pseudopotentials},
  author={Hamann, DR},
  journal={Physical Review B},
  volume={88},
  number={8},
  pages={085117},
  year={2013},
  publisher={APS}
}

@article{schlipf2015optimization,
  title={Optimization Algorithm for the Generation of ONCV Pseudopotentials},
  author={Schlipf, Martin and Gygi, Fran{\c{c}}ois},
  journal={Computer Physics Communications},
  volume={196},
  pages={36--44},
  year={2015},
  publisher={Elsevier}
}

@article{perdew1996generalized,
  title={Generalized Gradient Approximation Made Simple},
  author={Perdew, John P and Burke, Kieron and Ernzerhof, Matthias},
  journal={Physical Review Letters},
  volume={77},
  number={18},
  pages={3865},
  year={1996},
  publisher={APS}
}

@article{liu1989limited,
  title={On the Limited Memory BFGS Method for Large Scale Optimization},
  author={Liu, Dong C and Nocedal, Jorge},
  journal={Mathematical Programming},
  volume={45},
  number={1-3},
  pages={503--528},
  year={1989},
  publisher={Springer}
}

@article{methfessel1989high,
  title={High-precision sampling for Brillouin-zone integration in metals},
  author={Methfessel, MPAT and Paxton, AT},
  journal={physical review B},
  volume={40},
  number={6},
  pages={3616},
  year={1989},
  publisher={APS}
}

@article{nambu1960quasi,
  title={Quasi-Particles and Gauge Invariance in the Theory of Superconductivity},
  author={Nambu, Yoichiro},
  journal={Physical Review},
  volume={117},
  number={3},
  pages={648},
  year={1960},
  publisher={APS}
}

@article{pinsook2024analytic,
  title={Analytic solutions of Eliashberg gap equations at superconducting critical temperature},
  author={Pinsook, Udomsilp and Natkunlaphat, Nattawut and Rientong, Komkrit and Tasee, Pakin and Seeyangnok, Jakkapat},
  journal={Physica Scripta},
  volume={99},
  number={6},
  pages={065211},
  year={2024},
  publisher={IOP Publishing}
}

@article{noffsinger2010epw,
  title={EPW: A Program for Calculating the Electron–Phonon Coupling Using Maximally Localized Wannier Functions},
  author={Noffsinger, Jesse and Giustino, Feliciano and Malone, Brad D and Park, Cheol-Hwan and Louie, Steven G and Cohen, Marvin L},
  journal={Computer Physics Communications},
  volume={181},
  number={12},
  pages={2140--2148},
  year={2010},
  publisher={Elsevier}
}

@article{ponce2016epw,
  title={EPW: Electron–Phonon Coupling, Transport and Superconducting Properties Using Maximally Localized Wannier Functions},
  author={Ponc{\'e}, Samuel and Margine, Elena R and Verdi, Carla and Giustino, Feliciano},
  journal={Computer Physics Communications},
  volume={209},
  pages={116--133},
  year={2016},
  publisher={Elsevier}
}

@article{giustino2007electron,
  title={Electron–Phonon Interaction Using Wannier Functions},
  author={Giustino, Feliciano and Cohen, Marvin L and Louie, Steven G},
  journal={Physical Review B},
  volume={76},
  number={16},
  pages={165108},
  year={2007},
  publisher={APS}
}

@article{margine2014two,
  title={Two-Gap Superconductivity in Heavily n-Doped Graphene: Ab Initio Migdal–Eliashberg Theory},
  author={Margine, ER and Giustino, Feliciano},
  journal={Physical Review B},
  volume={90},
  number={1},
  pages={014518},
  year={2014},
  publisher={APS}
}

@article{zhao2019two,
  title={Two-Gap and Three-Gap Superconductivity in AlB$_2$-Based Films},
  author={Zhao, Yinchang and Lian, Chao and Zeng, Shuming and Dai, Zhenhong and Meng, Sheng and Ni, Jun},
  journal={Physical Review B},
  volume={100},
  number={9},
  pages={094516},
  year={2019},
  publisher={APS}
}

@article{gao2020strong,
  title={Strong-Coupling Superconductivity in LiB$_2$C$_2$ Trilayer Films},
  author={Gao, Miao and Yan, Xun-Wang and Lu, Zhong-Yi and Xiang, Tao},
  journal={Physical Review B},
  volume={101},
  number={9},
  pages={094501},
  year={2020},
  publisher={APS}
}

@article{modak2021prediction,
  title={Prediction of Superconductivity at 70 K in a Pristine Monolayer of LiBC},
  author={Modak, P and Verma, Ashok K and Mishra, Ajay K},
  journal={Physical Review B},
  volume={104},
  number={5},
  pages={054504},
  year={2021},
  publisher={APS}
}

@article{seeyang2025phase_japtwpgap,
  title={Phase Stability and Superconductivity in Hydrogenated and Lithiated Janus GaXS$_2$ (X = Ga, In) Monolayers},
  author={Seeyangnok, Jakkapat and Pinsook, Udomsilp},
  journal={Journal of Applied Physics},
  volume={138},
  number={16},
  year={2025},
  publisher={AIP Publishing}
}

@article{xie2024strong,
  title={Strong Electron–Phonon Coupling and Multigap Superconductivity in 2H/1T Janus MoSLi Monolayer},
  author={Xie, Hongmei and Huang, Zhijing and Zhao, Yinchang and Huang, Hao and Li, Geng and Gu, Zonglin and Zeng, Shuming},
  journal={The Journal of Chemical Physics},
  volume={160},
  number={23},
  year={2024},
  publisher={AIP Publishing}
}

@article{seeyangnok2026tunable,
  title={Tunable Superconductivity in Functionalized Janus MoSeA (A= H, Li) Monolayers: Competition between Lattice Instability and Electron Pairing},
  author={Seeyangnok, Jakkapat and Pinsook, Udomsilp and Ackland, Graeme J},
  journal={ACS Applied Energy Materials},
  year={2026},
  publisher={ACS Publications}
}

@article{liu2022two,
  title={Two-Gap Superconductivity in a Janus MoSH Monolayer},
  author={Liu, Peng-Fei and Zheng, Feipeng and Li, Jingyu and Si, Jian-Guo and Wei, Liuming and Zhang, Junrong and Wang, Bao-Tian},
  journal={Physical Review B},
  volume={105},
  number={24},
  pages={245420},
  year={2022},
  publisher={APS}
}

@article{seeyangnok2026stability,
  title={Stability, Electronic Disruption, and Anisotropic Superconductivity of Hydrogenated Trilayer Metal Tetraborides (MB4H; M= Be, Mg, Ca, Al)},
  author={Seeyangnok, Jakkapat and Ackland, Graeme J and Pinsook, Udomsilp},
  journal={Advanced Theory and Simulations},
  volume={9},
  number={3},
  pages={e70356},
  year={2026},
  publisher={Wiley Online Library}
}

@article{geim2013van,
  title={Van der Waals heterostructures},
  author={Geim, Andre K and Grigorieva, Irina V},
  journal={Nature},
  volume={499},
  number={7459},
  pages={419--425},
  year={2013},
  publisher={Nature Publishing Group UK London}
}

@article{jseeyang_ti2csh,
  title = {Theoretical prediction of structural stability and superconductivity in Janus Ti${}_{2}$CSH MXene},
  author = {Seeyangnok, Jakkapat and Pinsook, Udomsilp},
  journal = {Phys. Rev. B},
  pages = {--},
  year = {2026},
  month = {Jan},
  publisher = {American Physical Society},
  doi = {10.1103/dxg1-bxj4},
  url = {https://link.aps.org/doi/10.1103/dxg1-bxj4}
}

@article{seeyangnok2024superconductivity,
  title={Superconductivity and Electron Self-Energy in Tungsten-Sulfur-Hydride Monolayer},
  author={Seeyangnok, Jakkapat and Ul Hassan, M Munib and Pinsook, Udomsilp and Ackland, Graeme},
  journal={2D Materials},
  volume={11},
  number={2},
  pages={025020},
  year={2024},
  publisher={IOP Publishing}
}

@article{seeyangnok2024superconductivitywseh,
  title={Superconductivity and Strain-Enhanced Phase Stability of Janus Tungsten Chalcogenide Hydride Monolayers},
  author={Seeyangnok, Jakkapat and Pinsook, Udomsilp and Ackland, Graeme J},
  journal={Physical Review B},
  volume={110},
  number={19},
  pages={195408},
  year={2024},
  publisher={APS}
}

@article{qiao2024prediction,
  title={Prediction of Charge Density Wave, Superconductivity and Topology Properties in Two-Dimensional Janus 2H/1T-WXH (X = S, Se)},
  author={Qiao, Shu-Xiang and Jiang, Kai-Yue and Sui, Chang-Hao and Xiao, Peng-Cheng and Jiao, Na and Lu, Hong-Yan and Zhang, Ping},
  journal={Materials Today Physics},
  volume={46},
  pages={101485},
  year={2024},
  publisher={Elsevier}
}

@article{gan2024hydrogenation,
  title={Hydrogenation-Induced Superconductivity in Monolayer},
  author={Gan, Geng-Run and Fu, Si-Lie and Wang, Chun-An and Xie, Ya-Peng and Gao, Xue-Lian and Wang, Lin-Han and Chen, Yu-Lin and Chen, Jia-Ying},
  journal={EPL},
  volume={145},
  number={5},
  pages={56002},
  year={2024},
  publisher={EDP Sciences, IOP Publishing and Societ{\`a} Italiana di Fisica}
}

@article{fu2024superconductivity,
  title={Superconductivity in the Janus WSH Monolayer},
  author={Fu, Si-Lie and Gan, Geng-Run and Wang, Chun-An and Xie, Ya-Peng and Gao, Xue-Lian and Wang, Lin-Han and Chen, Yu-Lin and Chen, Jia-Ying and Wu, Xian-Qiu},
  journal={Journal of Superconductivity and Novel Magnetism},
  pages={1--9},
  year={2024},
  publisher={Springer}
}

@article{seeyangnok2025competition,
  title={Competition Between Superconductivity and Ferromagnetism in 2D Janus MXH (M = Ti, Zr, Hf; X = S, Se, Te) Monolayer},
  author={Seeyangnok, Jakkapat and Pinsook, Udomsilp and Ackland, Graeme J},
  journal={Journal of Alloys and Compounds},
  volume={1033},
  pages={180900},
  year={2025},
  publisher={Elsevier}
}

@article{seeyangnok2026triangular,
  title={Triangular Charge-Density Waves (T-CDW) Stabilize Janus Group-VI Chalcogenide Hydrides},
  author={Seeyangnok, Jakkapat and Pinsook, Udomsilp and Ackland, Graeme J},
  journal={arXiv preprint arXiv:2606.04954},
  year={2026}
}

@article{seeyangnok2026electron,
  title={Electron-Phonon Coupling and Charge Density Wave Instabilities in W2N and Halogen-Functionalized W2N Monolayers},
  author={Seeyangnok, Jakkapat and Pinsook, Udomsilp},
  journal={arXiv preprint arXiv:2606.04953},
  year={2026}
}

@article{seeyangnok2026enhanced,
  title={Enhanced and Tunable Superconductivity Enabled by Mechanically Stable Halogen-Functionalized Mo2C MXenes},
  author={Seeyangnok, Jakkapat and Pinsook, Udomsilp},
  journal={arXiv preprint arXiv:2602.11552},
  year={2026}
}

\end{document}